\documentclass[
11pt,
superscriptaddress,
showkeys,
showpacs,
nofootinbib
]{revtex4-1}

\usepackage[utf8]{inputenc}
\usepackage[english]{babel}
\usepackage{amsmath}
\usepackage{amsfonts}
\usepackage{amssymb}
\usepackage{graphicx}

\usepackage{color}
  
\newcommand{\Name}[1]{{\scshape #1},}
\newcommand{\Review}[1]{{\itshape #1},}

\newcommand{\Vol}[1]{{\bfseries #1}}

\newcommand{\Year}[1]{(#1)}

\newcommand{\Page}[1]{{\normalfont #1}}

\newcommand{\Pages}[2]{{\normalfont #1-#2}}
\newcommand{\Book}[1]{{\itshape #1}}
\newcommand{\Publ}[1]{{\normalfont(#1)}}

\newcommand{\URL}[1]{url: \url{#1}} 

\renewcommand{\div}{\mathrm{div} \, } 
\newcommand\rot{\mathrm{rot} \,} 

\newcommand{\bfA}{{\mathbf{A}}}

\newcommand{\bfE}{{\mathbf{E}}}

\newcommand{\bfH}{{\mathbf{H}}}

\newcommand{\bfe}{{\mathbf{e}}}

\newcommand{\bfr}{{\mathbf{r}}}

\newcommand{\rmc}{{\mathrm c}}
\newcommand{\rmd}{{\mathrm d}}
\newcommand{\upd}{{\, \mathrm d}}
\newcommand{\rme}{{\mathrm e}}

\newcommand{\rmg}{{\mathrm g}}
\newcommand{\rmi}{{\mathrm i}}

\newcommand{\cS}{{\mathcal S}}
\newcommand{\cV}{{\mathcal V}}

\newcommand{\CC}{{\mathbb C}}

\newcommand{\NN}{{\mathbb N}}

\newcommand{\ZZ}{{\mathbb Z}}

\newcommand{\sA}{{\mathsf A}}
\newcommand{\bsA}{{\boldsymbol{\mathsf{A}}}}

\newcommand{\sC}{{\mathsf C}}

\newcommand{\sE}{{\mathsf E}}
\newcommand{\bsE}{{\boldsymbol{\mathsf{E}}}}

\newcommand{\bsF}{{\boldsymbol{\mathsf{F}}}}

\newcommand{\bsH}{{\boldsymbol{\mathsf{H}}}}
\newcommand{\sI}{{\mathsf I}}

\newcommand{\sK}{{\mathsf K}}

\newcommand{\sV}{{\mathsf V}}

\newcommand{\frK}{{\mathfrak{K}}}

\newcommand{\frP}{{\mathfrak{P}}}

\newcommand\melec{m}

\newcommand\abs[1]{{| #1 |}}

\begin{document}

\title{Electromagnetic power and momentum in $N$-body hamiltonian approach to wave-particle dynamics in a periodic structure}

\author{Damien~F.~G.~Minenna}%
 \email[Electronic address: ]{damien.minenna@univ-amu.fr}
\affiliation{%
Centre National d'\'Etudes Spatiales, 31401 Toulouse cedex 9, France
}%
\affiliation{%
Aix-Marseille University, UMR 7345 CNRS PIIM, \'equipe turbulence plasma, case 322 campus Saint J\'er\^ome,  av.\ esc.\ Normandie-Niemen, 13397 Marseille cedex 20, France
}%
\affiliation{%
Thales Electron Devices, rue Lat\'eco\`ere, 2, 78140 V\'elizy, France
}%
\author{Yves~Elskens}%
 \email[Electronic address: ]{yves.elskens@univ-amu.fr}
\affiliation{%
Aix-Marseille University, UMR 7345 CNRS PIIM, \'equipe turbulence plasma, case 322 campus Saint J\'er\^ome,  av.\ esc.\ Normandie-Niemen, 13397 Marseille cedex 20, France
}%
\author{Fr{\'e}d{\'e}ric~Andr{\'e}}%
 \email[Electronic address: ]{frederic.andre@thalesgroup.com}
\affiliation{%
Thales Electron Devices, rue Lat\'eco\`ere, 2, 78140 V\'elizy, France
}%
\author{Fabrice~Doveil}%
\affiliation{%
Aix-Marseille University, UMR 7345 CNRS PIIM, \'equipe turbulence plasma, case 322 campus Saint J\'er\^ome,  av.\ esc.\ Normandie-Niemen, 13397 Marseille cedex 20, France
}%

\date{March 29, 2018. Submitted. \copyright The authors}

\begin{abstract}
To model momentum exchange in nonlinear wave-particle interaction, as in amplification devices like traveling-wave tubes, we use an $N$-body self-consistent hamiltonian description based on Kuznetsov's discrete model, and we provide new formulations for the electromagnetic power and the conserved momentum. This approach leads to fast and accurate numerical simulations in time domain and in one dimensional space.
\end{abstract}

    
\maketitle

The electromagnetic power and its density are major characteristics of amplifying devices 
and there is an increasing need to monitor them accurately in nonlinear regimes. 
While this power is generally computed using harmonic models, we undertake to compute it in position and time.
To model the wave-particle interaction, we choose the particle description (a.k.a.\ discrete particle or $N$-body description), 
combined with a self-consistent hamiltonian~\cite{myn78,ten94,els03}, 
because nonlinear regimes are well addressed by considering the evolution of $N$ nearly resonant charged particles interacting with the waves. 
Indeed, to predict turbulences and instabilities (due in particular to the chaotic dynamics of electrons) of the wave-particle system, 
the $N$-body description is the most intuitive approach. 
But it is also the most complex one in electrodynamics 
since the number of degrees of freedom for fields and particles involved is absolutely enormous.
We propose a solution to this problem.

\bigskip

The interaction between an electron beam and electromagnetic waves 
is one of the most fundamental processes in the physics of hot and cold, 
natural, industrial and laboratory plasmas. 
This process is also at the heart of state-of-the-art wave amplifiers like 
vacuum electronic tubes (traveling-wave tubes (TWTs), klystrons, etc.), 
free electron lasers or particle accelerators. 
The key mechanism in this interaction is the exchange of momentum, 
as e.g.\ in the Landau effect~\cite{lan46,one71}, 
when the phase velocity of the wave is close to the particles speed~\cite{dov05,dov18}. 
Nowadays, the large power and broad band spectrum used in those electronic devices lead to critical instabilities and are difficult to simulate.
For example, spurious frequencies are generated very far from the drive frequency in the nonlinear regime because of drive-induced oscillations. 

To follow particle motions and wave propagation in this interaction, 
we need better tools to study saturated nonlinear regimes, where electron motion perturbations become significant.
Currently, two options are available to model such system, each with typical drawbacks.  
The first option is kinetic models in time domain such as particle-in-cell (PIC) algorithms based on Vlasov equations 
(like e.g.\ \textsc{cst-particle} 
and \textsc{mafia}) \cite{CST,MAFIA}. 
But the resulting codes are extremely slow due to their large number of degrees of freedom. 
Therefore, they are not suitable for design activities and lead industry to rely on the second option~: 
specialized algorithms in frequency domain. 
Those specialized models (like e.g.\ \textsc{mvtrad}, \textsc{bwis} and \textsc{christine}) \cite{wal99,li09,An97} 
are industrial standards because they are fast and rigorous in their definition domains, 
but they are generally not suited for nonlinear regimes (instability saturations) 
and have difficulties at predicting oscillation phenomena due to their nature. 
Besides, they are not always fully consistent with Maxwell equations \cite{the16}.

We propose a third option, using a many-body description to design a specialized model in time domain. 
By definition, this approach may involve a priori far more degrees of freedom 
than grid-based or spectral discretisations of the kinetic one. 
But a field decomposition provided by the Kuznetsov discrete model \cite{kuz80} 
allows us to drastically reduce the number of simulation parameters, 
for fields and particles as well. 

The main purpose of this letter is to investigate the electromagnetic power distribution (see eq.~\eqref{e.Pn} below) in the Kuznetsov discretization.
Indeed, an important feature of traveling-wave tubes (TWTs) is the interaction efficiency (ratio between output power of the wave and electric power used by the device).
Combining this model with the hamiltonian formalism~\cite{and13} leads to a better respect of conservation properties in the model,
including Poincar\'e-Cartan invariants~\cite{arn89}, thanks to adapted symplectic algorithms~\cite{hai03,hai10}, 
allowing one to increase the numerical time step without incurring too much error on results.
With the hamiltonian formalism, we compute the momentum balance (see eq.~\eqref{e.NoeMom} below) 
in the wave-electrons interaction for a periodic structure, and design a numerical integrator.
Finally, simulations for a typical TWT will be run in one space dimension (1D), 
to be compared with a frequency algorithm to assess the accuracy of our time-domain approach.

\bigskip

This letter rests on ref.~\cite{and13}, which presents the construction of an electromagnetic hamiltonian, 
starting with the Kuznetsov discrete model \cite{kuz80,rys09,ber11}. 
Given the wave\-guide structure with period $d$ along the $z$-axis, with period (cell) index $n$, 
this model provides an exact discretized decomposition of the radiofrequency (RF), divergence-free electromagnetic fields 
in the cell-based representation,
\begin{align}
  \bfE(\bfr,t) 
  & = \sum_{s,n} \sV^s_{n}(t) \, \bsE^s_{-n}(\bfr) \, , 
  \label{e.VsnEsn} 
  \\
  \bfH(\bfr,t) 
  & = \rmi \sum_{s,n} \sI^s_{n}(t) \, \bsH^s_{-n}(\bfr) \, , 
  \label{e.IsnHsn}
\end{align}
where basis fields $\bsE^s_{n}$, $\bsH^s_{n}$ are obtained (see \cite{and13}) from Gel'fand transforms \cite{gel50}, 
and where $s \in \NN$ is the label for modes of propagation. 
Those basis fields depend only on the geometry of the wave\-guide.
They do not obey orthogonality conditions in a simple cell~; 
they decay (with oscillations) as $\abs{n - z/d} \to \infty$, 
but do not plainly vanish for $\abs{n - z/d} > 1/2$. 
The $\rmi$ factor in \eqref{e.IsnHsn} is a convenient choice,
for which $\sV^s_{n}$, and $\sI^s_{n}$ are real-valued.
In presence of electric sources, time-dependent coefficients $\sV^s_{n}$ and $\sI^s_{n}$ do not coincide, otherwise the system will violate Maxwell equations.
Note that it is possible to use a linearised version of the discrete model \cite{ter17,min17b}.

An advantage of this decomposition is that, 
for each mode of propagation, there are $2 n_{\mathrm{max}}$ different time variables (viz.\ $n_{\mathrm{max}}$ degrees of freedom) 
for fields in a wave\-guide of $n_{\mathrm{max}}$ periods. 
In comparison, finite difference techniques used in particle-in-cell codes necessitate millions of degrees of freedom to reach the same accuracy. 
Besides, this decomposition is exact regardless of the structure geometry provided it is periodic along the propagation direction.

Since terms $\sV^s_{n}$ and $\sI^s_{n}$ will be obtained from the hamitonian dynamics, we must express the electromagnetic power using the discrete model. This is done below.

\bigskip

In the Kuznetsov discrete model, the $n$-based representation derives from a $\beta$-based representation\footnote{In 
  contrast with refs \cite{rys09,and13}, we let the phase per pitch, $\beta d$, range over $[- \pi, \pi]$ 
  instead of $[0, 2 \pi]$, to enhance symmetry in calculations.}  
\begin{equation}
  \bfE(\bfr,t) = (2 \pi)^{-1} \sum_s \int^{\pi}_{-\pi} \sV^s_{\beta}(t) \, \bsE^s_{\beta}(\bfr) \, \rmd (\beta d)  \, ,
  \label{e.EfieldBetaRepre} 
\end{equation}
and similarly for the magnetic field with $\sI^s_{\beta}(t)$ and $\rmi \bsH^s_{\beta}(\bfr)$, 
where  $\sV^s_{\beta}, \sI^s_{\beta}$, 
$\bsE^s_{\beta}, \bsH^s_{\beta}$ 
are Fourier transforms of $\sV^s_{n}, \sI^s_{n}, \bsE^s_{n}$ and $\bsH^s_{n}$. 
These (propagating) modes are eigenvectors of the Helmholtz equation \cite{kuz80,rys09,and13}, 
with eigenvalues $\Omega^s_{\beta}$, 
\begin{align}
  \rot \bsE^s_{\beta}(\bfr) 
  &= - \rmi \mu_0 \Omega^s_{\beta} \bsH^s_{\beta}(\bfr) \, , 
  \label{e:Helmo1} 
  \\
  \rot \bsH^s_{\beta}(\bfr) 
  &= \rmi \epsilon_0 \Omega^s_{\beta} \bsE^s_{\beta}(\bfr) \, , 
  \label{e:Helmo2}
\end{align}
for solenoidal eigenfields meeting the boundary conditions on the wave\-guide wall 
and the Floquet condition $\bsE(\bfr + n d \bfe_z) = \rme^{-\rmi n \beta d} \bsE(\bfr)$, 
so that the propagation constant $\beta$ is the wavenumber associated with Bloch's theorem.

We normalize these eigenfields to
\begin{equation}
  N^s_{\beta} \delta^s_{s'} 
  = \int_{\cV_{0}} \epsilon_0 \bsE^s_{\beta} \cdot \bsE^{s'*}_{\beta} \upd^3 \bfr 
  = \int_{\cV_{0}} \mu_0 \bsH^s_{\beta} \cdot \bsH^{s'*}_{\beta} \upd^3 \bfr \, ,
  \label{e.Omesb}
\end{equation}
corresponding to the electric or magnetic energies stored in one period over the cell-volume $\cV_0$. 
For later convenience, this normalisation is chosen\footnote{As in \cite{and13}, we choose $N^s_{\beta}=\Omega^s_{\beta}$ which has the dimension of a pulsation, 
   and then $\sV^s_{n}, \sI^s_{n}$ are homogenous to the square root of an action. 
   In \cite{rys09}, the normalisation has the dimension of an energy and $\sV^s_{n}$ and $\sI^s_{n}$ become dimensionless.} 
equal to the eigenvalues $\Omega^s_{\beta}$. 
The Kronecker symbol $\delta^s_{s'}$, expressing orthogonality of modes $s \neq s'$, 
follows from the nondegeneracy of eigenvalues ($\Omega^{s}_\beta \neq \Omega^{s'}_\beta$), 
which furthermore forces the integral 
$\int_{\cV_0} ( \bsE^{s}_{\beta} \wedge \bsH^{s' *}_{\beta} \big) \cdot \bfe_z \, \rmd^3 \bfr$
to vanish, where $\bfe_z$ is the $z$-axis unit vector. 

Derivating \eqref{e.Omesb} with respect to $\beta$, 
and using derivatives of Helmholtz eqs \eqref{e:Helmo1} and \eqref{e:Helmo2} and the derivative of the Floquet condition, 
yield the group velocity $v_{\rmg \, \beta}^s = \partial_{\beta} \Omega^s_{\beta}$ of the electromagnetic wave along the $z$-axis,
\begin{equation} \label{e.vgS}  
  v_{\rmg \, \beta}^s  = \frac{d}{\Omega^s_{\beta}} \int_{\cS} \Re \big( \bsE^s_{\beta} \wedge \bsH^{s*}_{\beta} \big) \cdot \bfe_z \, \rmd^2 \bfr \, ,
\end{equation}
where $\cS$ is the transverse section of the wave\-guide. 

%
%
 
\bigskip

Now, we recall the hamiltonian approach~\cite{and13}. 
The Poynting energy $H_{\textrm{em}}= \frac{1}{2} \int_{\cV_{\ZZ}} (\epsilon_0 |\bfE|^2 + \mu_0 |\bfH|^2) \, \rmd^3 \bfr$ 
over the system volume $\cV_{\ZZ}$, 
with normalisation \eqref{e.Omesb} and the Parseval relation, 
yields the hamiltonian for radiative fields
\begin{equation}
\label{e.Hem}
  H_{\textrm{em}} 
  = \frac{1}{2} \sum_{s} \sum_{n_1, n_2} \Big( \sV^s_{n_1} \Omega^s_{n_1-n_2} \sV^s_{n_2} + \sI^s_{n_1} \Omega^s_{n_1-n_2} \sI^s_{n_2} \Big) \, .
\end{equation}
The normalisation \eqref{e.Omesb} ensures that $\sV^s_{n}$ and $\sI^s_{n}$ 
are canonical variables\footnote{Instead
   of these ``cartesian'' variables for oscillators, one can use the angle-action approach 
   by considering $\sqrt{2} \sC^s_{n}(t) = \sV^s_{n}(t) + \rmi \sI^s_{n}(t) \in \CC$, 
   where canonical actions are $\sC^s_{n} \sC^{s*}_{n}$, with conjugate angles $\theta^s_{n} = {\mathrm{Arg}}\, \sC^s_n$. 
   Then the electromagnetic hamiltonian \eqref{e.Hem} becomes 
   \begin{equation}
      H_{\textrm{em}} 
      = \sum_{s} \sum_{n_1, n_2} \sC^s_{n_1} \sC^{s*}_{n_2} \, \Omega^s_{n_1-n_2} \, .
   \end{equation}
   }, 
satisfying the Poisson brackets $\{ \sV^s_{n_1},\sV^{s'}_{n_2} \} = \{ \sI^s_{n_1},\sI^{s'}_{n_2} \} = 0$ 
and $\{ \sV^s_{n_1},\sI^{s'}_{n_2} \} = \delta^{n_1}_{n_2} \delta^{s}_{s'}$, 
with generalized coordinates $\sI^s_{n}$ and conjugate momenta $\sV^s_{n}$.  

The Fourier transform of $\Omega^s_{\beta}$, denoted $\Omega^s_{n_1-n_2}$, 
appears in \eqref{e.Hem} as the coupling coefficient between ``RF oscillators'' (or coupled cavities) at cells $n_1$ and $n_2$. 
For coupled-cavity traveling-wave tubes, the actual dispersion relation is well described 
as a nearest-neighbour oscillator coupling, viz.\ with $\Omega_n = 0$ for $| n | \geq 2$ \cite{rys09}, 
whereas for helix traveling-wave tubes the coupling is longer-ranged (typically, $\Omega_n > 0$ for $|n| \leq 7$) \cite{the16,min17a}. 
This difference in the coupling coefficients reflects the properties of the underlying basis fields, 
which in turn reflect the actual geometry of the slow wave structure. 


The coupling with particles involves the longitudinal $N$-body dynamics \cite{els03,and13}. 
We consider a beam of $N$ electrons with mass $\melec$ and charge $-|e|$, labelled $k$, 
with positions $z_k$ and canonical momenta $p_k$.
The magnetic potential 
of the wave is rewritten with the cell-based representation~\eqref{e.IsnHsn}
as $\bfA(\bfr,t) = \rmi  \sum\limits_{sn} \sI^s_{n}(t) \bsA^s_{-n}(\bfr)$ 
(with Coulomb gauge $\div \bfA = 0$), 
where the basis magnetic potential $\bsA^s_{n}$ is imaginary-valued\footnote{In ref.~\cite{and13},
   we write $\bsF^s_{n} = - \rmi \bfA^s_{n}$. Note typos in ref.~\cite{and13}~: 
   its eq.~(15) misses a minus sign in $\rme^{- \rmi n \beta d}$, 
   and a $c^2$ factor is missing in the second member of the second equation of its eq.~(26).} 
like $\bsH^s_{n}$.
The electrons ballistic motion and their coupling with fields  
is given in one dimension by the relativistic hamiltonian\footnote{The
   non-relativistic version of \eqref{e.Hcpl} is
   \begin{equation}
      H_{\mathrm{el}}= \sum^N_{k=1} \frac{\melec \dot{z}^2_k}{2} + |e|  \sum_{s,n} \sum^N_{k=1}  \dot{z}_k \sI^s_{n}(t) \rmi \sA^s_{-n}(z_k) \, .
   \end{equation}
}
\begin{equation}
  H_{\mathrm{el}}  = \sum^N_{k=1} \melec c^2 \, \Big( \big[1- \frac{\dot{z}^2_k}{c^2} \big]^{-1/2} -1 \Big) = \sum^N_{k=1} \melec c^2(\gamma_k -1) \, , 
  \label{e.Hcpl} 
\end{equation}
with the Lorentz factor
\begin{equation}
  \gamma_k = \bigg[ 1 +  \Big( p_k + |e| \sum_{s,n} \rmi \sI^s_{n} \sA^s_{-n}(z_k) \Big)^2 \Big/ ( \melec c )^2 \bigg]^{1/2} \, .
\end{equation}
The Coulomb interaction in the beam (possibly taking into account the waveguide wall boundary conditions) 
is incorporated through a scalar potential $\phi$.
Then the Poynting energy splits into an RF contribution \eqref{e.Hem}, and a space charge energy 
\begin{equation} 
  H_{\mathrm{sc}}  
  = - \frac{1}{2} \sum^N_{k=1} \sum_{k' \neq k} |e| \, \phi(z_k - z_{k'}) \, , 
  \label{e.Hsc}
\end{equation}
with the coulombian field $E_{\mathrm{sc},z} = - \partial_{z} \phi$ involving all charged particles of the system to interact with each other.
The sum of \eqref{e.Hem}, \eqref{e.Hcpl} and \eqref{e.Hsc} is the total self-consistent hamiltonian 
for wave-particle dynamics along the $z$-axis. 

\bigskip

In this formulation, the system total momentum 
is the sum of electron (mechanical) and field (Poynting, a.k.a.~Abraham) 
momenta 
\begin{equation}
  \mathsf{P}_{\mathrm{wp},z} = \sum^N_{k=1} \gamma_k \melec \dot{z}_k
        + \sum_{s}\sum_{n_1, n_2} \sV^s_{n_1} \sI^s_{n_2} \sK^s_{n_1-n_2} \, ,
  \label{e.NoeMom}
\end{equation}  
where the geometric kernel $\sK^s_{n}$ is constructed from 
\begin{equation}
  \sK^s_{\beta} 
  = c^{-2} \, v_{\rmg \, \beta}^s \, \Omega^s_{\beta} \, , 
  \label{e:Ksbeta}
\end{equation} 
and the group velocity  $v_{\rmg \, \beta}^s = \rmi \sum_n n d \, \Omega^s_{n} \, \rme^{\rmi n \beta d}$.
We introduce the modified (by the $\rmi$ factor) inverse Fourier transform of \eqref{e:Ksbeta} 
\begin{align}
  \sK^s_{n} & = (2\pi)^{-1} \Re \Big[ \int^{\pi}_{-\pi} \rmi \, \sK^s_{\beta} \, \rme^{- \rmi n \beta d}  \, \upd(\beta d) \Big] \\
  & = c^{-2} d \sum_{n_2} (n_2 - n) \, \Omega^s_{n-n_2} \, \Omega^s_{n_2} \, ,
  \label{e:Ksn}
\end{align}
where $\sK^s_{n} = - \sK^s_{-n}$.

The Legendre transform of the total hamiltonian is the lagrangian, for which
Noether's theorem shows \eqref{e.NoeMom} to be  
conserved during the wave-electron momentum exchange.
However, applying Noether's theorem is not trivial, 
as the wave\-guide geometry is not invariant under infinitesimal translations,  
but the lagrangian is invariant under 
$z_k \mapsto z_k + \varepsilon, \sI_\beta \mapsto \rme^{- \rmi \beta \varepsilon} \sI_\beta$.
Of course, the hamiltonian, being time-independent, is also the conserved total energy, 
though the energy does not split into a mere sum of individual, single-particle and single-mode contributions. 
Note that the wave-particle interaction is based on momentum exchange \cite{els03,dov05}, 
as implied by \eqref{e.NoeMom}. 
The conservation of this momentum is essential to ensure that simulations are consistent.

\bigskip

The power of the RF electromagnetic wave is a key feature of vacuum electron devices. The fields representation \eqref{e.VsnEsn}-\eqref{e.IsnHsn} provides an expression for the power,
obtained from the flux of the Poynting vector,
$\frP_z (z,t) = \int_{\cS}(\bfE \wedge \bfH)\cdot \bfe_z \, \rmd \cS$, 
\begin{equation}
  \frP_z(z,t) 
  = \sum_{s,s'} \sum_{n_1, n_2} \, \sV^{s}_{n_1}(t) \, \sI^{s'}_{n_2}(t) \, \frK^{s,s'}_{n_1, n_2}(z) \, ,
  \label{e.PVIK}
\end{equation}
with
\begin{equation} \label{e:ksn1n2}
\frK^{s, s'}_{n_1, n_2}(z)
   =   \int_\cS \Big( \bsE^{s}_{-n_1}(\bfr) \wedge \rmi \bsH^{s'}_{-n_2}(\bfr) \Big) \cdot \bfe_z \, \rmd \cS \, .
\end{equation}
With this representation, the longitudinal total electromagnetic power 
$P_{\mathrm{tot},z}(t) = \int^{\infty}_{-\infty} \frP_z (z, t) \, \rmd z$
reads 
\begin{equation}
  P_{\mathrm{tot},z}(t) 
  = (2 \pi)^{-1} \sum_s \int^{\pi}_{-\pi} \sV^s_{\beta}(t) \, \rmi \sI^s_{\beta}(t) \, \sK^s_{\beta} \, c^2  \, \rmd (\beta d)  \, ,
  \label{e:Pzbeta}
\end{equation}
This yields the cell-based representation of the total electromagnetic power 
\begin{equation}
  P_{\mathrm{tot},z}(t) 
  =  c^2 \sum_s \sum_{n_1, n_2} \sV^s_{n_1}(t) \, \sI^s_{n_2}(t) \, \sK^s_{n_1-n_2} \, ,
  \label{e.Ptotz}
\end{equation}
which depends only on time. 

The contribution of cell $n$ to the electromagnetic field power \eqref{e.PVIK} is
\begin{equation}
  P_{z, n} (t)  = \sum_{s, s'} \sum_{n_1, n_2} \sV^{s}_{n_1} \, \sI^{s'}_{n_2} \, 
              \int_{(n- \frac{1}{2})d}^{(n+\frac{1}{2})d}  \frK^{s, s'}_{ n_1, n_2} (z) \, \rmd z \, .
  \label{e.PnK}
\end{equation}
These time-domain expressions are suited for observing, for example, wave oscillations and transient reflections on the power, 
which is impossible from frequency models.
To obtain the contribution of each cell, we expand \eqref{e.PnK} by considering the $z$-dependence in $\frK^{s,s'}_{n_1, n_2}$. 
First, we note that, in $\beta$-representation, \eqref{e.PnK} involves two wave numbers $\beta_1$ and $\beta_2$ 
(from the Fourier transform on $n_1$ and $n_2$).  
In contrast with \eqref{e.Ptotz}, we cannot impose $\beta_1 = \beta_2$ by invoking destructive interference between different wavenumbers, 
because we consider only a finite cell $(n-1/2) d < z < (n+1/2) d$, not the full line $-\infty < z < \infty$.

To compute $\frK^{s, s'}_{n_1, n_2}$, 
we need the $z$-dependence of the basis fields $\bsE^s_{n_1}, \bsH^{s'}_{n_2}$,
or equivalently of eigenfields $\bsE^s_{\beta_1}, \bsH^{s'}_{\beta_2}$.
The main difficulty in \eqref{e.PnK} lies in the eigenmode cross-term \eqref{e:ksn1n2}. 
For $\beta_1 = \beta_2$, it relates directly with the group velocity \eqref{e.vgS}. 
As a rule of thumb, because $(\beta_1 - \beta_2) d$ is not very large
(in the so-called continuous waveform (CW) regime of prime interest to applications),
the simple approximation
\begin{equation} \label{e:kapprox}
  \int_{(n- \frac{1}{2})d}^{(n+\frac{1}{2})d}  \frK^{s, s'}_{ n_1, n_2} (z) \, \rmd z
  \approx 
  c^2 \sK^s_{n_1-n_2} \, \frac{1}{2} \, (\delta_{n_1}^n + \delta_{n_2}^n) \, \delta_{s}^{s'}  
\end{equation}
yields the estimated power
in cell $n$ 
\begin{equation} 
  P_{z,n}(t)  
  \approx 
 c^2 \sum_s \sum_{n_2} \frac{1}{2} \, (\sV^s_{n} \, \sI^s_{n_2} - \sV^s_{n_2} \, \sI^s_{n} ) \, \sK^s_{n-n_2} \, ,
  \label{e.Pn}
\end{equation}
which is time dependent, and where the geometric kernel $\sK^s_{n}$ depends on the dispersion relation $\Omega^s_{\beta}$ and not on transverse components of the field \eqref{e:ksn1n2}. 
Of course, the sum over $n$ of all \eqref{e.Pn} reduces to the total electromagnetic power \eqref{e.Ptotz}.

\bigskip

\begin{figure}
  \begin{center}
  \includegraphics[width=\columnwidth]{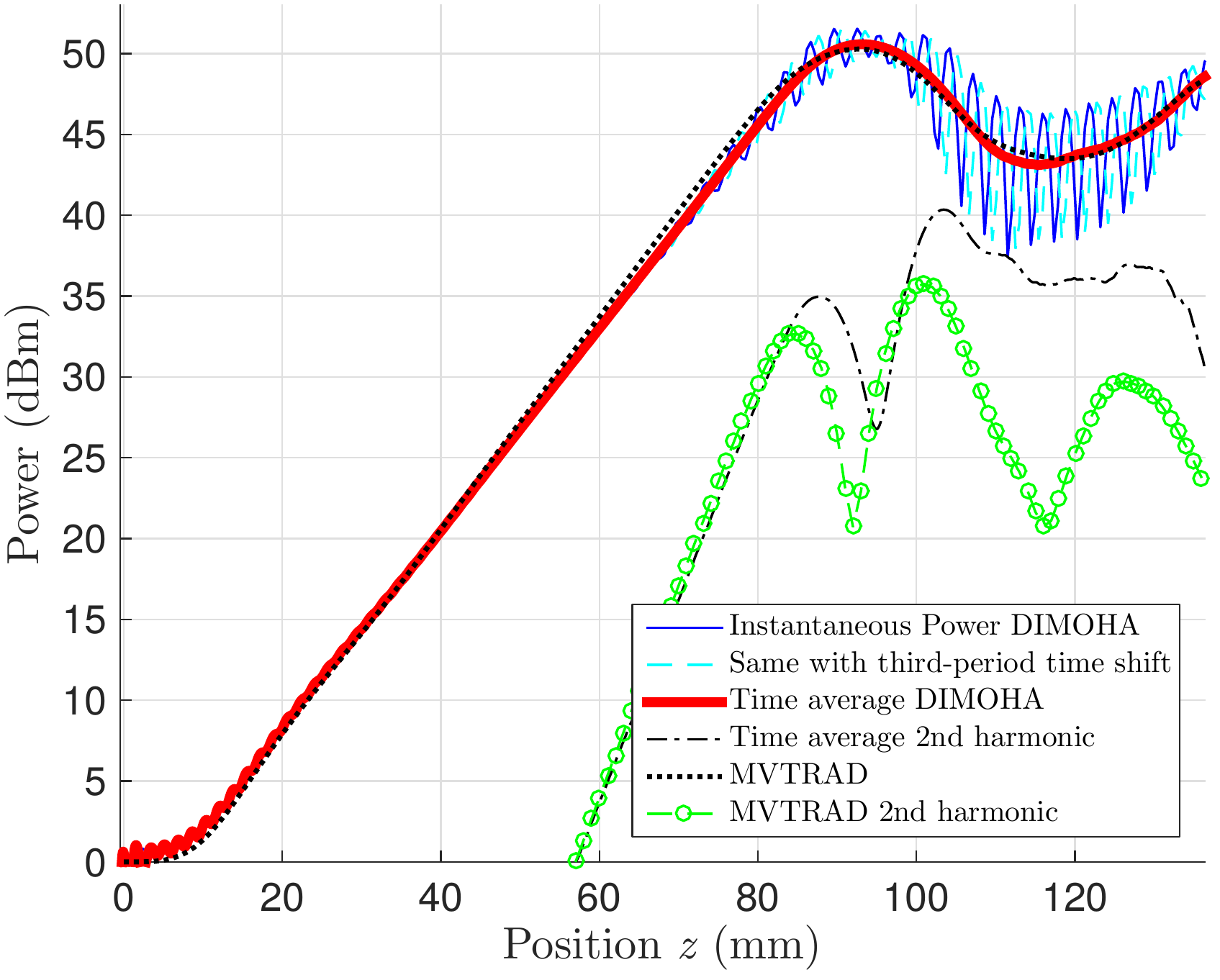}
  \caption{Spatial amplification of the electromagnetic power 
  inside a model traveling-wave tube (no taper or attenuations) for the fundamental mode ($F=12$~GHz) 
  and its second harmonic from 1D time-domain simulation {\sc{dimoha}} and frequency-domain algorithm {\sc mvtrad}. 
  Continuous blue (resp.\ cyan) curve~: instantaneous power \eqref{e.Pn} at $t_{\mathrm{final}} = 6$~ns 
  (resp.\ at $t_{\mathrm{final}} + 1/(3 F)$).
  Thick red curve~: power from \eqref{e.PE2}, 
  using for $\tilde E_z$ the time Fourier transform of the electric field \eqref{e.VsnEsn} (i.e.\ field envelope). 
  Black dashed curve~: its second harmonic. 
  Black (resp.\ green) dots~: fundamental (resp.\ second harmonic) power from {\sc mvtrad}.}
  \label{f.power}
  \end{center}
\end{figure}
  
To build an evolution algorithm from this formulation, 
we need the basis fields $\bsE^s_{n}$ for \eqref{e.VsnEsn}. 
In this letter, we consider only one mode of propagation (we omit superscript $s=0$) 
and the thin beam approximation (axial 1D beam). 
This system is classically described by means of an equivalent circuit model~\cite{pie50,and10,min17b} 
with a telegraph equation for the RF field, coupled with the beam. 
In the harmonic regime, the RF field evolves as $E_z(\bfr, t) = \tilde{E}_z(\bfr)  \mathrm{e}^{\rmi \omega t}$
and generates locally the (time-averaged) power crossing section $\cS$ at abscissa $z$,
\begin{equation} 
   \langle \frP_{z} \rangle 
   = \frac{| \tilde{E}_z (r = 0)|^2}{2 \beta^2  \,Z_{\rmc\, \beta}} \, ,
  \label{e.PE2}
\end{equation}
with the wave impedance (a.k.a.\ circuit impedance) $Z_{\rmc\, \beta}$.
As the angular frequency $\omega$ selects a single wavenumber $\beta$, 
the electric field on axis must also read $\sV_\beta(t) \sE_{\beta,z}(z)$,
with eigenfield component $\sE_{\beta,z}(z)$ and amplitude $\sV_{\beta}(t) = \tilde{\sV}_{\beta} \mathrm{e}^{\rmi \omega t}$.
And since the impedance $Z_{\rmc\, \beta}$ is defined independently of any beam,
\eqref{e.PE2} holds also without beam (``cold'' regime of the device), 
with evolution equations reducing to $\tilde{\sV}_{\beta} = \rmi \tilde{\sI}_{\beta}$. 
Then the time-averaged flux of the Poynting vector \eqref{e.PVIK} reduces to 
\begin{equation}
  \langle \frP_{z} \rangle 
  = \frac{1}{2}  \tilde{\sV}^{*}_{\beta} \, \rmi \tilde{\sI}_{\beta} \frac{v_{\rmg \, \beta} \, \Omega_{\beta}}{d}   \, .
  \label{e.avPz1}
\end{equation} 
Finally, the longitudinal component of the electric field on axis is proportional to $\rme^{- \rmi \beta z}$. 
So the electric eigenfield shape function is\footnote{Note 
   a difference by a $\sqrt{2}$ factor in \cite{ber11} due to missing the $1/2$ prefactor in the harmonic regime
   (the usual factor for average power using peak amplitudes) in front of \eqref{e.avPz1}. 
   Numerical comparisons with experimental data  
   validate our relation without the $\sqrt{2}$ factor.}
\begin{equation}
  \tilde{\sE}_{z,\beta} (z)
  = \rme^{- \rmi \beta z} \sqrt{\frac{v_{\rmg \, \beta} \, \Omega_{\beta}}{d} \, \beta^2 \, Z_{\rmc\, \beta}} \, .
  \label{e.EsbZ}
\end{equation}
The basis fields $\bsE^s_n$ in 1D are given by the Fourier transform of \eqref{e.EsbZ}, 
and are similar to cardinal sine functions. In \cite{and13}, we also have $\rmi \bsA^s_{\beta} = \bsE^s_{\beta} / \Omega^s_{\beta}$, providing the magnetic potential eigenfield $\bsA^s_n$.
Note that \eqref{e.EsbZ} involves only experimentally known values. 
The Fourier transform of \eqref{e.EsbZ} gives the $z$-projected $\sE_{n}$ to be used in \eqref{e.VsnEsn}. 
To obtain the actual electric field in our time-domain modeling, for arbitrary waves and in presence of a beam,
we only need to find coefficients $\sV_{n}(t)$. 

\bigskip

To benchmark our model, we run a one-dimensional numerical simulation 
with only the principal mode of propagation ($s = 0$). 
The three hamiltonian terms \eqref{e.Hem}, \eqref{e.Hcpl} and \eqref{e.Hsc} 
generate simple evolution equations amenable to explicit discrete-time symplectic maps \cite{arn89,hai03,hai10} 
for $\sI_n$, $\sV_n$, $z_k$ and $p_k$, providing an order 2 symmetric symplectic integrator. 
Physical model inputs are the dispersion relation $\Omega_{\beta}$ and impedance $Z_{\rmc\, \beta}$, 
which can be obtained experimentally, or from a Helmholtz solver (given the structure geometry).
For later comparisons, we take the dispersion relation used in \cite{min17b} for a generic traveling-wave tube.
These data yield the cell-based representation coefficients $\Omega_{n}$ and $\sK_{n}$, 
and field shape function $\sE_{n}(z)$.

We simulate the beam as a line of $N$ macro-electrons 
spaced from each other by $\delta z_{\mathrm{par}} = 1 \cdot 10^{-5}$~m, 
with identical initial speed $\dot{z}_{k,\mathrm{ini}} = 4.56 \cdot 10^7$~m/s 
and charge $Q = 103~270~|e|$ fixed by cathode potential and current values. 
The on-axis space-charge field in one dimension is found using Rowe's approximation \cite{ber11}.
Fields are discretized on a grid with mesh $\delta z_{\mathrm{field}}$. 
Time-sinusoidal waves are excited at the first cell by adding a forcing term $2 \pi F U \sin (2 \pi F t)$ to $\dot \sV_1$ 
in the integrator, 
with frequency $F = 12$~GHz, and appropriate amplitude $U$ to ensure\footnote{To characterize the gain, 
   in Fig.~\ref{f.power} we plot the power logarithmic ratio $10 \log_{10} ( \frP_z(z) / \frP_z(0) )$, 
   which defines the dBm scale in electronics.}
$\frP_z(z=0) = 1$~mW. 
Note that generation of second harmonic occurs in TWTs.
To represent the wave\-guide beginning and end on the $z$-axis, 
we broke conservation properties by appending long attenuators on either side 
to damp $\sV_{n'} (n' < 0 {\textrm{ or }} n' >n_{\mathrm{end}})$ at a small rate in space.

The comparison is performed versus algorithm {\sc mvtrad} \cite{wal99} (property of Thales Electron Devices), 
an industrial code in frequency domain (or so-called envelope model), 
well-used and robust to simulate industrial traveling-wave tubes (TWTs) in the GHz regime 
but known to have some difficulty at estimating accurately effects in nonlinear regimes and at predicting oscillation phenomena. 
One reason for these discrepancies may be that frequency models are not fully consistent with Maxwell equations \cite{the16}.

Since the aim of this paper is mostly to present theoretical tools for wave-particle simulations rather than to design a complete algorithm for scientific and industrial purposes, 
we focus on a single numerical comparison. 
Benchmarking with simulations and experimental data over wide ranges will be done in a separate work, 
with our algorithm modified to account for current TWT defects (like losses and attenuations or tapering)~: 
in the present letter, we focus on the fundamental physical issue of momentum exchange.

Fig.~\ref{f.power} presents a comparison between the electromagnetic power 
computed with our approach ({\sc dimoha} for HAmiltonian DIscrete MOdel) and {\sc mvtrad}. 
Initial parameters (like cathode current and potential, coupling impedances, phase velocity, and tube length) 
are set to ensure that the amplification leads the power to saturation before the end of the tube, 
and space charge effects are taken into account.
Our instantaneous power \eqref{e.Pn} at a time $t_{\mathrm{final}} = 6$~ns, is plotted (continuous thin blue curve). The cyan curve is also the instantaneous power \eqref{e.Pn}, but at the time $t_{\mathrm{final}} + 1/(3 F)$, and demonstrates that  \eqref{e.Pn} follows the wave propagation with time.
The thick red curve is the power from \eqref{e.PE2}, where we take the time Fourier transform over one time-period of the electric field \eqref{e.VsnEsn}, and the black dashed curve is its second harmonic.

Nonlinear effects (trapping) start to occur approximately after $z = 60$~mm when observing the electron dynamics (not shown in this letter, but presented in \cite{min17a}). 
The power saturation occurs at $z = 90$~mm, when trapped particles start to regain momentum.
Agreement between the frequency aspect of the discrete model, viz.\ time-averaged power \eqref{e.PE2}, 
and {\sc mvtrad} (black dots) is excellent for the fundamental mode
(actually, the discrepancy hardly catches the eye). 
Oscillations for the instantaneous time power \eqref{e.Pn} occur in nonlinear regime, 
probably caused by the approximation \eqref{e:kapprox}. 
Small variations between our approach and {\sc mvtrad} (green circles), such as a bias in saturation power, appear for the power's second harmonic. 
But second harmonic experimental measurements are difficult to perform accurately. Both our model and {\sc mvtrad} remain inside experimental uncertainty.

\bigskip

In this work, starting from a total hamiltonian describing the coupling 
between an electron beam and the radiative fields propagating in a periodic waveguide, 
and using a field discretization,
we constructed the conserved momentum of the system and instantaneous and time-averaged electromagnetic power. 
We built a time-domain algorithm which is much faster than industrial PIC codes (running in the scale of seconds) 
and as accurate as a frequency-domain algorithm in tested regimes. 
But this approach goes further than frequency models, as it enables to address situations in nonlinear regimes with trapped and chaotic particles interacting with fields, 
which is of interest to a broader community of physicists and engineers (like \cite{dov06}).

Oscillations, reflections and multitone operation are currently investigated. 
We are also working on two- and three-dimensional versions of the model 
to provide a complete electron velocity and position distribution.
To validate our approach, comparison with experimental TWTs will be performed (see \cite{min17a} for preliminary results), after taking into account losses and tube defects.

\section*{Acknowledgments}

The authors gratefully acknowledge fruitful comments from D.~Escande.

\end{document}